# Contribution of the electron-phonon interaction to Lindhard energy partition at low energy in Ge and Si detectors for astroparticle physics applications

Ionel Lazanu

University of Bucharest, Faculty of Physics, POBox MG-7, Bucharest-Magurele, Romania

Sorina Lazanu[*]

National Institute for Materials Physics, Str. Atomistilor 105bis, Magurele Ilfov, 077125, Romania

**Abstract**

The influence of the transient thermal effects on the partition of the energy of selfrecoils in germanium and silicon into energy eventually given to electrons and to atomic recoils respectively is studied. The transient effects are treated in the frame of the thermal spike model, which considers the electronic and atomic subsystems coupled through the electron – phonon interaction. For low energies of selfrecoils, we show that the corrections to the energy partition curves due to the energy exchange during the transient processes modify the Lindhard predictions. These effects depend on the initial temperature of the target material, as the energies exchanged between electronic and lattice subsystems have different signs for temperatures lower and higher than about 15 K. Many of the experimental data reported in the literature support the model.

**Keywords:** direct dark matter detection, nuclear recoil, low energy, ionisation, transient thermal effects

**PACS:** 29.40.-n Radiation detectors

**Highlights:**

- Correction to the Lindhard curves of energy partition for low energy selfrecoils based on the exchange of energy during transient thermal processes between electronic and atomic subsystems

- The correction is evaluated for Ge and Si selfrecoils and could improve the present limits of signals to be used in detection.

- Low and high temperature limits are calculated for the correction.

- A detailed and exhaustive analysis of some properties of Ge and Si, related to transient thermal effects, is performed.

[*] Corresponding author. Tel: +40213690185; Fax: +40213690177;
e-mail: lazanu@infim.ro

# Contribution of the electron-phonon interaction to Lindhard energy partition at low energy in Ge and Si detectors for astroparticle physics applications

## 1. Introduction

A major direction of investigation in experiments in astroparticle and particle physics is the direct detection of dark matter, as well as the test of hypotheses related to the existence of new types of elementary particles.

Experimental efforts of progressively reducing or rejecting background events have the aim of obtaining a spectrum of rare nuclear recoil events. The discrepancies observed between different measurements in detectors, in particular those based on germanium and silicon materials, during the last fifty years and covering a very large range of temperatures, as well as possible discrepancies with theoretical models put in evidence the importance of understanding the phenomena produced by low energy nuclear recoils. Thus the new experiments planned to be developed will be able to improve the present limits of signals to be used in detection.

Generally, an ion moving inside a medium loses its energy by collisions with electrons and nuclei (or atoms as a whole). Most detectors are sensitive only to the electronic energy loss used for ionization and (in some situations) to scintillation. Thus the knowledge of the energy loss sharing between processes is mandatory for a detailed understanding of the response to different particles and energies.

Formalisms describing the energy loss were developed based on the Thomas-Fermi model [1, 2] by Bohr [3], Bethe [4], etc. The most general formalism describing the partition of the energy of an incoming particle into a solid target was proposed by Lindhard and co-workers [5]: average effects are calculated, as result of the study of the division of the total incident energy into energy given to recoiling atoms and energy given to electrons. The image for the processes considered, in the frame of Lindhard's theory, is that the interaction between the incident particle and the solid proceeds as follows: the particle, heavier than the electron, with or without electrical charge, interacts with the electrons and with the nuclei from the lattice structure. It loses its energy in processes which depend on the nature of the particle and on its energy. The effect of the interaction of the incident particle with the electrons of the target is ionisation. The nuclear interaction produces mainly bulk defects. As a result of the interaction, depending on the energy and on the nature of the incident particle, the target nucleus could or could not be fragmented and/or removed from its lattice site. So, a recoil nucleus (nuclei) and in some cases one or more light particles are formed. The nucleus has charge and mass numbers lower or equal to that of the target. Then, the primary knock-on nucleus, if its kinetic energy (resulting from the energy transferred in the scattering process) is large enough, can produce the displacement of a new nucleus, and the process continues as long as the energy of the recoiling nucleus is higher than the threshold for atomic displacements. In its turn, the recoil nucleus loses energy both in ionization processes and in displacement ones. This phenomenon can be regarded as a cascade process.

Under a series of approximations, Lindhard and co-workers deduced an integro-differential equation for the energy deposited in ionization, $E_i$ (or correspondingly spent in atomic motion, $E_{ni}$) as a function of the kinetic energy of the incoming particle. These approximations can be formulated as: i) the recoil atoms with appreciable energies are not produced by electrons; ii) atomic binding energy term could be neglected; iii) the energies transferred to electrons in every scattering process are small compared with the projectile energy; iv) the separation of nuclear and electronic collisions is possible; v) the energy transferred in nuclear collision is small compared to the energy of the incoming particle.

A supplementary contribution to the energy partition, $\Delta E$ is given by the energy exchanged between the systems of electrons and atoms during the transient processes which follow the deposition of energy, and it will be calculated in the present paper. This is equivalent with the elimination of hypotheses iv) and v) of Lindhard theory. The scattering processes following the primary interaction of the incoming particle, as described in Lindhard's theory, are rapid processes, and the following them the electronic and atomic subsystems have different temperatures and evolve slowly during thermalisation processes. The coupling of the subsystems eventually produces a redistribution of the energy imparted to the target, and only the final state is accessible from experimental point of view. It is characterised by a modified energy partition: $E_i' = E_i \pm \Delta E$ and $E_{ni}' = E_i \mp \Delta E$. Due to the temperature dependence of the specific heats and thermal conductivities of the electronic and lattice systems, and of the strength of electron - phonon coupling, the effect of the transient processes on energy partition between the two subsystems depends also on the initial temperature of the target and to the peculiarities of the material (e.g. doping).

The data points measured by different experiments for the energy stored as ionization (and visible energy - scintillation when this phenomenon is possible) as a function on the energy of the recoil do not entirely agree neither with each other, nor with the curves calculated in the frame of Lindhard's theory.

In this paper, the contribution of transient processes is investigated in detail for two classical semiconductors: germanium and silicon, materials of interest for detectors, particularly for cryogenic detectors. A special attention is paid to the case of very low energy transferred in the interaction process, of interest for direct WIMPs detection for example, or interaction of low energy neutrinos. In these conditions, the original curves calculated in Lindhard theory are the initial step, and the new results obtained represent a correction to them.

After a compilation of data related to the partition of the selfrecoil energy in Ge and Si, we shall present succinctly the main hypotheses of the thermal spike model, in the frame of which transient processes are analysed. As the results obtained for the temperature distributions in space and time, for both electronic and atomic subsystems, depend on their specific heats and thermal conductivities, as well as on their coupling strength, a survey of literature data for these quantities in Ge and Si is presented. The energy transfer between the two subsystems during transient processes is then evaluated, and the consequences on Lindhard curves analysed.

## 2. Survey of data in the literature for Ge and Si

The data reported for the energy partition of selfrecoils in Ge and Si were measured using different methods, at different temperatures from room temperature (RT) up to cryogenic temperatures, on a time span of fifty years. Compilations of published data for Ge and Si are presented in Figs. 1 and 2 respectively. In the last years, the interest of the scientific community moved toward lower recoil energies, less than 1 keV.

As can be seen from Fig. 1, for energies of the selfrecoil less than 2 keV experimental data show a deviation from the Lindhard curves so that the energy eventually given to the electronic system is higher than that predicted by Lindhard theory. The data used in the compilation are from Refs. [6, 7, 8, 9, 10, 11, 12, 13, 14, 15, 16], and the dotted curve is the original Lindhard curve corresponding to the value $k$=0.1577 [5].

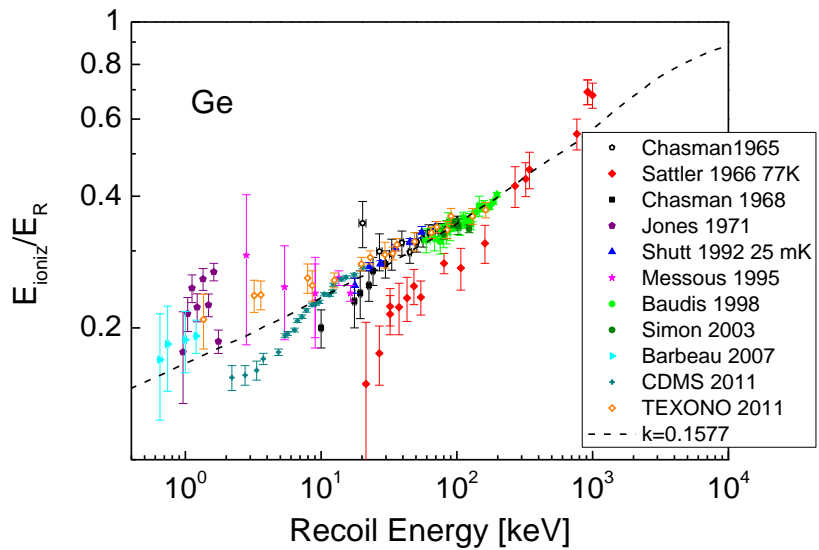

Fig. 1: Partition of selfrecoil energy in Ge

The compilation for Si, represented in Fig. 2, presents the same trend for selfrecoil energies lower than 10 keV. The data are from Refs. [17, 18, 19, 20]. The value of the parameter $k$ for Si is 0.1463. The corresponding Lindhard curve is also represented in the figure, as a dashed line. Most of the data seem to correspond to a curve situated at lower values of the ratio $E_{ioniz}/E_R$.

The major problem in the interpretation of the data for both Ge and Si is the lack of complete information on the properties of the material (as doping concentration and, correlated, concentration of free charge carriers) and also measurement temperature. When available, measurement temperatures were indicated in the legends of the graphs.

In order to explain the discrepancies between the measured and the calculated points, one must first revisit the calculations. A major problem is the accuracy of electronic and nuclear stopping powers for selfrecoils of low energy. In this energy range, the nuclear stopping power based on Thomas-Fermi potential is overestimated (see [21] and also sputtering data). Tilinin [22] argued that for low energies the electronic stopping power decreases more rapidly than the square root of the energy, fact which implies that a factor of up to 3 in respect to the curves calculated in the frame of Lindhard theory could appear. Arista and co-workers showed the impact parameter dependence of the electronic stopping power [23, 24]. To increase the agreement between theoretical calculations and experimental data some authors proposed to consider Lindhard curves corresponding to modified values for *k* parameter [16, 25].

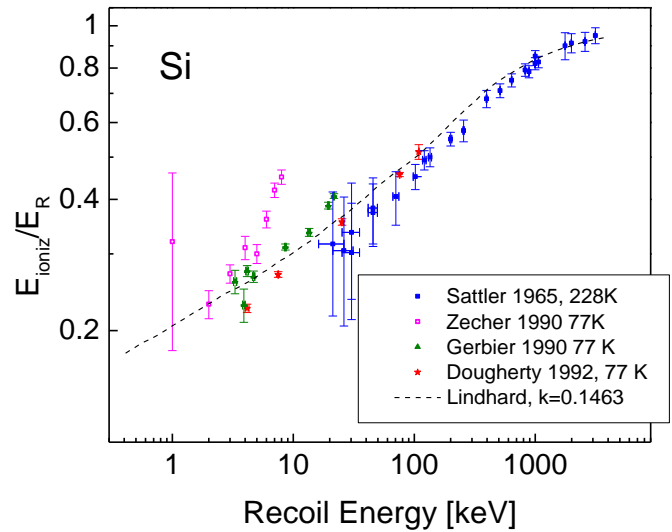

Fig. 2: Partition of selfrecoil energy in Si

Another possibility to explain the discrepancies, which will be investigated in the present paper, is to take care on the transfer of energy between the subsystems in the solid, i.e. electronic and atomic, during the transient thermal phenomena which follow the acts of interaction of an incoming particle in the target and the consequence of which is the slowing down.

## 3. Transient phenomena

Transient thermal processes following heavy ion irradiation are due to the deposition of an appreciable amount of energy in a small region of a material, and in a very short time interval. The spreading of this energy can be modelled based on the understanding of the mechanisms governing the coupling of electronic excitations to the damage produced in the lattice. There are more suggested models accounting on track formation due to swift heavy ions in materials [26]. One of them is the thermal spike. In this model, there are two subsystems, namely electrons and lattice, which are coupled through electron-phonon interaction. In the case of laser irradiation the heat is transferred to the electronic subsystem, while in heavy ion irradiation, depending on the ion and on its kinetic energy, the energy could be transferred in both directions. The thermal spike model was developed first for swift heavy ions, where electronic energy loss is predominant, and where hot electrons transfer their energy by electron-phonon coupling to the cold lattice. Heat diffusion in both subsystems is described by the classical heat equation, with the source given by the energy released by electronic energy loss (or by the laser beam), i.e. by a term *A(r,t)*. This energy transfer was used for the creation of lattice damage in swift ion irradiated materials, which were put in evidence experimentally [27]. For ion irradiation there exists a regime in which the nuclear energy loss is also important, and in this case the energy transferred to the lattice subsystem is the source in the heat equation for the atomic temperature, *B(r,t)*. We developed a model for the regime of comparable electronic and nuclear energy losses, by considering the contributions from both electronic and nuclear sources in semiconductors [28], and we studied also the case of dominant nuclear contribution to heat [29]. This last regime corresponds to very low energies of recoils and appears at the end of their range. In the frame of the thermal spike model, one considers the sample divided in very thin layers, perpendicular to the direction of the recoil. The heat phenomena which develop are followed in each of these thin layers. The energy lost by the recoil in each layer is imparted between the electronic and lattice (atomic) subsystems of the target, in accordance with the stopping powers corresponding to these processes. Due to the two types of interaction, the two subsystems, electronic and atomic, have different temperatures and are coupled through a term that is a measure of the energy exchange, the electron - phonon coupling factor *g*. A cylindrical symmetry is usually considered, due to the straight trajectory of the projectiles which lose energy – see the discussion on the applicability of the cylindrical and spherical spikes in Ref. [29]. The exchange of energy between neighbouring layers, perpendicular to the trajectory of the ion, is neglected. Thus, due to the energy transfer from the projectile (recoil) toward electrons and nuclei respectively, localized regions of the medium characterized by departure from equilibrium, characterized by the temperatures $T_e$ (electronic subsystem) and $T_a$ (atomic one) appear, and they differ in time and space extension, due to the fact that the mechanisms of interaction and the kinematics are distinct. The energy exchange between these subsystems is described by a term given by the product $g(T_e^p - T_a^p)$, with *p* = 1 at RT, describing Newton's law of cooling. The typical characteristic times of the two subsystems differ by about two orders of magnitude.

The dependence of these temperatures on the distance to the track of the recoil, *r*, and on the time after its passage, *t*, are solutions of two coupled partial differential equations [30]:

$$C_e(T_e)\frac{\partial T_e}{\partial t} = \frac{1}{r}\frac{\partial}{\partial r}\left[rK_e(T_e)\frac{\partial T_e}{\partial r}\right] - g(T_e^p - T_a^p) + A(r,t)$$
$$C_a(T_a)\frac{\partial T_a}{\partial t} = \frac{1}{r}\frac{\partial}{\partial r}\left[rK_a(T_a)\frac{\partial T_a}{\partial r}\right] - g(T_a^p - T_e^p) + B(r,t)$$
(1)

where $T_{e(a)}$, $C_{e(a)}$, and $K_{e(a)}$ are respectively the temperatures, the specific heat, and the thermal conductivities of the electronic (index *e*) and atomic (lattice) subsystems (index *a*).

The sources satisfy the conservation laws:

$$\int_0^\infty dt \int_0^\infty 2\pi r A(r,t)\, dr = \left(\frac{dE}{dx}\right)_{el}$$
$$\int_0^\infty dt \int_0^\infty 2\pi r B(r,t)\, dr = \left(\frac{dE}{dx}\right)_{n}$$
(2)

with $\left(\frac{dE}{dx}\right)_{el}$ and $\left(\frac{dE}{dx}\right)_{n}$ the electronic and nuclear linear energy losses, respectively.

We would like to emphasize that in the model of the thermal spike the energy is stored in both subsystems in the form of heat, and consequently can be exchanged based only on the temperature difference. In fact, part of the energy in the atomic subsystem is stored in defect production, some of them being 'permanent', while the energy in the electronic subsystem is used for excitation and ionization and is imparted to electrons as kinetic energy, being taken away by delta electrons. Consequently, not all the energy is available to be exchanged, as the model supposes, and it is difficult to assess the quantity (or percentage) which might be extracted from the source terms. In a previous paper, the authors evaluated the amount of energy that is stored in defects at cryogenic temperatures [31].

## 4. Numerical results, analysis and interpretation

The solution of the system (1) depends strongly on the parameters of the electronic and atomic subsystem, on their temperature dependencies, as well as on the coupling parameter *g*. Not all these parameters are known from experiment at the temperatures of interest, as will be explained in the following.

### 4.1 Temperature dependences of specific heats and thermal conductivities of electronic and atomic subsystems of Si and Ge

a) **Specific heat**

The heat capacity is the amount of heat required to raise the temperature of an object or substance with one degree. Usually, heat capacity is divided by the amount of substance, mass, or volume, so defining molar heat capacity (heat capacity per mole), specific heat capacity (or specific heat as heat capacity per unit mass) and volume (volumetric) specific heat (as specific heat per unit volume).

The specific heat of a material is the sum of the contributions from the lattice and electronic systems.

In the frame of Debye theory, the low temperature limit ($T \ll \theta_D$) of the lattice specific heat of crystalline solids is represented by $T^3$ dependence; here $\theta_D$ stands for the Debye temperature, i.e. the temperature at which all normal modes of the lattice are excited. The low temperature limit of the Debye temperature is 645 K in Si and 374 K in Ge [32]. In agreement with Blackman's calculations [33], a $T^3$ dependence of the atomic specific heat is valid at temperatures below $\theta_D/50$, or at least $\theta_D/100$. In the high temperature limit, the Dulong-Petit law is valid, with a temperature independent specific heat.

There is little information about the electronic system in semiconductors. The information about it is extrapolated from metals. The electronic specific heat is proportional with the concentration of free electrons. In fact, in the measurement of the specific heat one measures the total specific heat, i.e. the sum of atomic and electronic components. These two components are comparable only in the very low temperature limit, while at higher temperatures the electronic specific heat is negligible in respect to the atomic one. Keesom and Pearlman showed that below 5K the specific heat capacity of both Si [34] and Ge [35] can be decomposed into a $T^3$ term corresponding to the contribution from the lattice (atomic system), and a linear one, corresponding to the electronic system:

$$C_p = \alpha T^3 + \gamma T \qquad (3)$$

In silicon, the measurements for the specific heat at constant pressure ($C_p$) reported by different authors [34, 36, 37, 38] agree quite well. In Figure 3, we represent the temperature dependence of the lattice volumetric specific heat of Si at constant pressure as experimental data, together with the fit. For temperatures below 2.5 K, the fit curve is extrapolated. In the calculations, we used the following dependences in different temperature ranges, obtained by fitting the experimental data, expressed in J/(cm$^3$K):

$$C_a = 2.05 \times 10^{-7} T_a^{3.37} \qquad T_a \leq 27.7 \text{ K}$$
$$C_a = -0.243 + 9.76 \times 10^{-2} T_a - 1.7 \times 10^{-5} T_a^2 + 1.40 \times 10^{-8} T_a^3 - 5.59 \times 10^{-12} T_a^4 + 8.46 \times 10^{-16} T_a^5 \qquad 27.7 < T_a \leq 1960$$

(4)

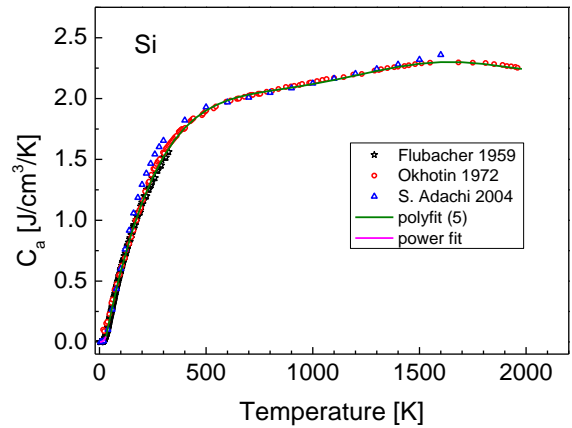

Fig. 3: Temperature dependence of the lattice volumetric specific heat of Si at constant pressure

For the lattice heat capacity of germanium, the data of Keesom and Pealman [35], Piesbergen [39] and Okhotin [37] were used. Using the data reported in the literature, we obtained the following dependences of $C_a$ (in J/cm³/K):

$$C_a = 5.85 \times 10^{-6} T_a^{3.11} \quad\quad T_a \leq 24.7 \text{ K}$$
$$C_a = -0.293 + 1.8 \times 10^{-1} T_a - 4.63 \times 10^{-5} T_a^2 + 1.27 \times 10^{-8} T_a^3 \quad\quad 24.7 < T_a \leq 189.5$$
$$C_a = 1.256 + 9 \times 10^{-4} T_a - 6.44 \times 10^{-7} T_a^2 + 1.61 \times 10^{-10} T_a^3 \quad\quad T_a > 189.5$$

(5)

The fit functions used in the calculations, together with the extrapolation at lower temperatures, are represented in Fig. 4 below:

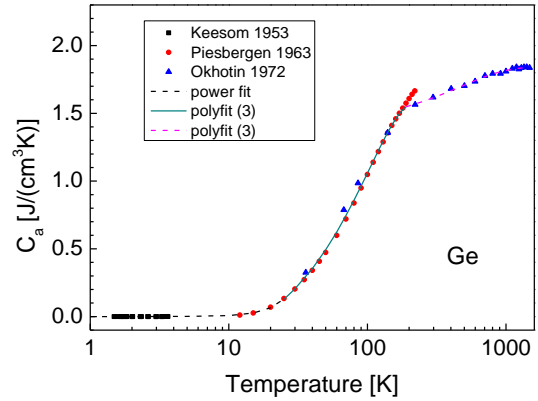

Fig. 4: Temperature dependence of the lattice specific heat of Ge at constant pressure

While the lattice specific heat is a characteristic of the material, the electronic one depends on the concentration of free electrons in the sample, i.e. in the equilibrium measurements it depends on the doping of the sample. In the range of cryogenic temperatures, for silicon, Pearlman and Keesom [34], Niinikoski [40], Wagner [41], Tsibidis [42] reported for the coefficient of the temperature γ from eq. (3) in the electronic specific heat values between $10^{-7}$ and $10^{-6}$ J/g/K² depending on the doping species and concentration. These values are measured in equilibrium conditions, and correspond to temperatures less than 1K. In Ref. [43], the authors estimate a ratio of electronic to atomic specific heats in non-degenerate semiconductors (following laser excitation) equal to the ratio of the concentrations of thermal carriers, which around RT is generally $10^{-9} - 10^{-4}$. In the case we are interested in, this ratio is even smaller, of the order 8x10⁻¹¹. In the literature regarding the thermal spike in semiconductors, different authors proposed different approaches for the electronic specific heat. So, in the study of InP irradiated with swift heavy ions, in Refs. [44, 45], $C_e$ was taken independent on temperature and having the value: $C_e = 3/2\, n_e k_B$, with $n_e$ the concentration of electrons freed due to the deposition of the electronic energy lost by the ion, calculated in Refs. [44, 45] to be in the order of 6x10²² cm⁻³, this way the resultant electronic specific heat being more orders of magnitude higher that the equilibrium one. Toulemonde *et al.* [46] take the thermodynamic parameters for the electron subsystem consistent with those ascribed previously by them to metallic materials [47] and insulators [48]. While the first case does not rise problems, in insulators the authors suppose that hot electrons excited in the conduction band by the passage of the ion behave like free electrons in metals, and supposing that the number of freed electrons per atom is about 1, and using the Dulong-Petit formula, an electronic specific heat constant in temperature, of about 1 J cm$^{-3}$ K$^{-1}$ is obtained. For Si, these authors used also a temperature independent specific heat. Subsequently, they refined the thermodynamic parameters of the electronic subsystem [49], considering both for metals (Au) and for insulators (SiO$_2$) electronic specific heats depending linearly on temperature up to a certain temperature which in metals is the Fermi temperature (the temperature corresponding to Fermi energy) and in dielectrics the temperature corresponding to the band gap ($T_g = E_g/k$), and then a constant value.

For germanium, the situation is qualitatively the same. Measurements of the electronic specific heat at low temperatures (less than 1 K) were reported by Bryant and Keesom [50], Aubourg [51], Wang [52], Olivieri [53]. Depending on the doping species and concentration, the temperature coefficient γ was found to be between $2 \times 10^{-7}$ and $4 \times 10^{-6}$ J/g/K$^2$; Keesom reported in [35] a value of γ which depends on the free carrier concentration $n$ as: $\gamma = 2.15 \times 10^{-11} \mu \cdot n^{2/3}$ (in J/mol/K$^2$), where $\mu$ is the ratio of effective carrier mass to free electron mass. In its turn, $\mu$ was found to be roughly independent on n and on temperature, having the value 0.25. At high temperature, the electronic specific heat attains a constant value.

**b) Thermal conductivity**

It is generally accepted that the carriers of the thermal energy in crystalline materials are phonons and charged carriers, and the thermal conductivity is additive:

$$K = K_L + K_e \tag{6}$$

where $K_L$ and $K_e$ are the thermal conductivity components due to the transport of heat by lattice waves and electrons respectively.

In semiconductors, the lattice conductivity is the most important in nearly the whole range of temperatures of interest, i.e. where the material is solid. The theory of lattice thermal conductivity was reviewed successively by Peierls [54], Ziman [55], and Carruthers [56]. The case of semiconductors is discussed by Drabble and Goldsmid [57]. In the frame of Debye theory, in the relaxation time approximation, lattice thermal conductivity is calculated as:

$$K_L(T) = \frac{k_B}{6\pi^2} \cdot \left(\frac{k_B T}{\hbar}\right)^2 \sum_{i=1}^{3} \frac{1}{v_i} \int_0^{\theta_D/T} \tau_i(x,T) \cdot \frac{x^4 e^x}{(e^x-1)^2} dx \tag{7}$$

where $x = \hbar\omega/(k_B T)$, $k_B$ in Boltzmann's constant, and the sum is over the three polarisations. The total relaxation time is found by summing the reciprocals of the relaxation times corresponding to different scattering mechanisms [58]:

$$\tau_i^{-1} = \tau_{B,i}^{-1} + \tau_I^{-1} + \tau_3^{-1} + \tau_4^{-1} + \tau_{ep,i}^{-1} \tag{8}$$

where $\tau_{B,i}$, $\tau_I$, $\tau_3$ ($\tau_N$, $\tau_U$), $\tau_4$, $\tau_{ep,i}$ are the relaxation times associated with phonon scattering on boundaries, isotopes, three phonon processes (normal and umklamp), four phonon ones, and neutral impurities respectively.

Therefore, the reciprocal of thermal conductivity is the sum of reciprocal conductivities corresponding to each of the mentioned mechanisms, which in turn are prevailing in different temperature ranges. These temperature ranges are usually quantified in terms of the Debye temperature of the crystal. So, at the lowest temperatures, where the wavelength of the lattice vibrations is relatively large, boundary scattering limits the conductivity [59], which has a $T^3$ dependence, as has the lattice specific heat, and in pure crystals attains a maximum at about $0.05 \times \theta_D$ [60]. Near the maximum, thermal conductivity is determined mainly by isotope scattering, and then it decreases with the increase of temperature due to normal (N) 3-phonon processes. At around $0.1 \times \theta_D$, $K_L$ decreases due to Umklapp (U) processes. The relaxation time corresponding to U scattering is proportional to $1/T$ [61]. At higher temperatures, in the range of temperatures in the order or higher than $\theta_D$ it becomes necessary to consider the relaxation times determined by 4 phonon processes, derived by Pomeranchuck [62]. Thermal conductivity decreases faster than $1/T$ at these temperatures [63].

All data reported in the literature reflect the type of temperature dependence of $K_L$ described above, but both the temperature corresponding to the maximum ($T_{max}$) and the value of the thermal conductivity ($K_{Lmax}$) depend on the analysed sample (sample size, doping, isotope content).

For silicon, as pointed out by Glassbrenner and Slack [63], the measurements reported by different authors agree quite well on the descending part. The differences between the values published, corresponding to very low temperatures, are due to boundary scattering in different size samples and to phonon scattering from impurities and other crystalline defects. The increase of purity leads to a lower value of $T_{max}$, and to a higher value of $K_{Lmax}$. The value of $T_{max}$ varies between 22.5 [64] and 200 K [65], while the corresponding $K_{Lmax}$ varies between 52 and 0.4 W/(cmK) respectively.

For germanium there is a similar situation. The value of $T_{max}$ is situated between 12.5 [66] and 70 K [67], and the corresponding $K_{Lmax}$ values are 18 and 0.6 W/(cmK) respectively.

In an attempt to standardise the data, the National Bureau of Standards of USA together with the American Institute of Physics proposed [71] for pure silicon that at low temperatures the "most probable curve" is based on Holland's [64] data, and below the lowest temperature available in his measurements the curve is extrapolated parallel to the data reported by Carruthers *et al*. [68], considering a $T^{2.9}$ dependence, up to 1 K. At 120 K the recommended curve merges into the curve made by the data of Glassbrenner and Slack [63], up to the melting point, as shown in Figure 5.

In our calculations, we consider below 1K a $T^3$ dependence of $K_L$, as shown by the extrapolation represented by the red dash-dotted curve, and at high temperatures the extrapolation proposed by Glassbrenner and Slack [63], represented by the blue dashed curve. The following dependences, obtained from fit, are to be used:

$$K_a = \begin{cases} 0.09T^3(0.016 * \exp[-0.05T] + \exp[-0.14T] & T \leq 120.7 \\ 13 \times 10^3 \times T^{-1.6} & T > 120.7 \end{cases} \quad (9)$$

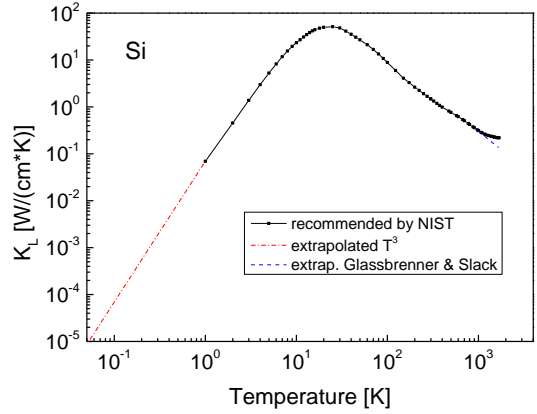

Figure 5: Lattice thermal conductivity of Si

The same publication (working group) proposed for germanium, for $K_L$, the utilisation of the data of Slack and Glassbrenner [66] in the range from 3 to 940 K. At this last temperature, the electronic polar and ambipolar thermal conductivities become important, and an extrapolation for $K_L$ is proposed by the authors, represented by the blue dashed curve in Fig. 6. For temperatures in the range 1-3 K the recommended curve is traced parallel to the data of Carruthers [68, 69] and of Geballe and Hull [70] up to 1 K. Below this temperature, we used a $T^3$ extrapolation, shown as the red dash-dotted curve.

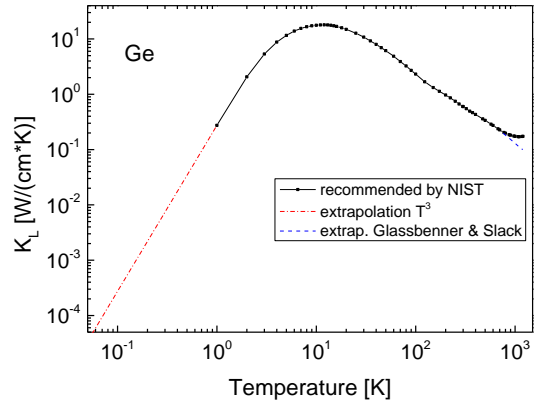

Figure 6: Lattice thermal conductivity of Ge

In what regards the electronic thermal conductivity, its importance appears especially in metals, and consequently the theory elaborated for metals [71] was first used in semiconductors, as in the case of the electronic specific heat. Ure [72, 73] devised a method to separate the measured thermal conductivity into its lattice and electronic components at temperatures higher than $\theta_D/2$. The method requires that the temperature is low enough so that ambipolar diffusion is not present. The value of the electronic thermal conductivity depends on the concentration of charge carriers, and increases with the doping of the semiconductor. At very low temperatures, the electronic thermal conductivity is proportional to the electronic heat capacity, i.e. is proportional with the temperature. It was shown theoretically [57, 74, 75] and put in evidence experimentally [63] for Si and Ge that the electronic thermal conductivity is the sum between a polar and a bipolar part: the polar $K_{ep}$ is the usual Wiedemann-Franz-Lorenz contribution known from metals, while the bipolar $K_{eb}$ is a property of semiconductors, caused by electron-hole pairs, with energy equal to the band gap, diffusing down the temperature gradient, and is significant at temperatures higher than Debye temperature. According to Glassbrenner and Slack [63], in solid silicon, around the melting point, 63 % of heat conduction is by phonons, 32% by electronic bipolar contribution, and the rest of 5% by the electronic polar contribution. In the evaluation of the electronic contribution, the electric conductivity of the material is very important, together with the dependence of the electric parameters on temperature (e.g. the semiconductor band-gap and the ratio of electron to hole mobilities). At cryogenic temperatures, $K_e$, like $C_e$, is proportional with the temperature, the proportionality factor depending on the concentration of free electrons. In the whole interval of variation of the temperature of interest, we used for the temperature dependence of $K_e$ the expression [63, 75]:

$$K_e(T) = \left[2 + \frac{b}{(1+b)^2}\left(\frac{E_g}{k_B T} + 4\right)^2\right]\left(\frac{k_B}{e}\right)^2 \sigma T \tag{10}$$

where $b$ is the ratio of electron to hole mobilities. $b$, $E_g$ and $\sigma$ are temperature dependent, the last one through the concentration of free carriers and through their mobilities.

### c) Electron – phonon coupling

The coupling of the electronic and atomic subsystems is studied both theoretically and experimentally [55, 76]. It has been found to be very important in explaining the structural modifications of the target material around the track of an energetic ion, which loses energy predominantly through ionization, i.e. is in the high electronic energy loss regime [77]. This coupling is the base of the thermal spike model, but also of the atomistic simulations. The electron-phonon coupling has been extensively studied in metals. Extending the arguments of Flynn and Averback [78], Finnis and co-workers developed a formula for the estimation of the electron-phonon coupling constant [79]. For the electron with a mean free path $\lambda_{mfp} = r_0 T_0 / T_a$, the electron-phonon coupling is:

$$g = \frac{k_B^2 \theta_D D(\varepsilon_F) v_F T_e}{r_0 T_0} \tag{11}$$

where $D(\varepsilon_F)$ is the electronic density of states at the Fermi level, $v_F$ the Fermi velocity, $r_0$ is Wigner-Seitz radius and $T_0$ is the atomic temperature at which the mean free path $\lambda_{mfp}$ equals $r_0$.

Kaganov *et al.* [80] found for the electron-phonon coupling factor the expression:

$$g = \frac{\pi m_e n_e v_s^2}{6 \tau_e T_e} \tag{12}$$

under the following approximations: the electronic system is treated as a free electron gas, and the electron-atom system is not far out of equilibrium ($|T_a - T_e| \ll T_a$). Here $n_e$ is the density of electrons, $v_s$ is the velocity of sound in the lattice, and $\tau_e = \lambda_{mfp}/v_F$. It has been shown [81] that the two expressions are equivalent.

In metals, both experimental and calculated values for the electron-phonon coupling factor are in the range 1 – 1000 x $10^{10}$ W/(cm³K$^{-1}$). In Ref. [82], by assuming that the electrons equilibration is dependent only on electron temperature, one finds that $\tau_e \sim 1/T_e$ so that g is temperature independent.

In the low temperature limit, the electron-phonon coupling term does not have the linear form of the Newton law of energy transfer: $g(T_a - T_p)$ and this is replaced by a term of the type $g(T_a^p - T_e^p)$. In pure metals, the electron-electron interaction time is much smaller than the electron-phonon one, so that $\tau_e$ from formula (12) is essentially $\tau_{e-ph}$, which in its turn is inversely proportional to the number of thermal phonons, and consequently $1/\tau_{e-ph} \propto T^3$ [83]. Therefore p=5 in the transfer term [84]. In disordered materials, including highly doped ones, and in the situation the phonon dimensionality differs from 3D, the relaxation time shows a different dependence on temperature, and the exponent p has been found to be 4 or 6 [84]. In these materials, two scattering processes are the most important: the electron-phonon scattering and the inelastic scattering on impurities. Usually, the dimensionless parameter $q\lambda_{mfp}$ is used, with $\lambda_{mfp} = v_F \tau_e$ and $q$ the momentum transferred to the electron due to electron phonon scattering. In the clean limit, this product is much greater than unity, and the contrary is valid for the dirty limit. For the last limit, $1/\tau_{e-ph} \propto T^2$ if the scattering centres are static (and p=4), and $1/\tau_{e-ph} \propto T^4$ if the impurities are vibrating ($p = 6$).

In applying the thermal spike model to semiconductors, the electron-phonon coupling constant is either evaluated using formula (12) [44, 45] or kept as a free parameter of the model [85]. Values in the order $10^{12} - 10^{13}$ W/(cm³K$^{-1}$) are generally used in calculations of thermal spike development in semiconductors irradiated at RT.

Recent studies related to bolometres for direct detection of dark matter [86], to MOSFETs [87] and SOI devices operating at sub-Kelvin temperatures [88], to metal–insulator superconducting junctions [89], put in evidence that the transfer term in the low temperature limit has a similar expression in semiconductors as in metals. The major difference in respect to metals, which have a spherical Fermi surface, is that the application of the free electron model is questionable. The Fermi surface in Si and Ge is six-fold degenerate and the intervalley scattering is the most important process in the relaxation of the momentum and energy of the electrons. The electron-phonon relaxation time depends on the velocity of sound on the direction of the *q* vector, and on its value as: $1/\tau_{e-ph} \propto v_s^{r+1} q^r$ [90], and after calculations one finds p=r+4. Sota *et al*. [91] have found from theory for 3D phonons in Si the exponent *r* = 2, and therefore *p* = 6. Different values for *p* have been measured: *p* = 6 [92], *p* = 5 [87]. In the literature, there are no reports related to the temperature extension of the coupling with *p*>1, the experimental measurement being limited at tenths of K; also, there are no mentions on the coupling coefficient at temperature lower than room temperature.

## 4.2 Numerical solutions for time and space dependencies of atomic and electron temperature, influence of the parameters of the electronic subsystem

As can be concluded from the analysis of the experimental data related to the specific heats and thermal conductivities of the electronic and atomic systems of Si and Ge, there is accurate knowledge of the parameters of the atomic system, while much uncertainty exists for those of the electronic system, and also for the coupling factor *g*. A new physics could be hidden behind these coefficients, at it was appreciated in the literature [93].

We investigated the influence of the parameters of the electronic system, specific heat and thermal conductivity dependence on electronic temperature, on the development of the thermal spike. So, a Si selfrecoil of 50 keV kinetic energy (with electronic and nuclear energy losses of 238.3 and 343.8 keV/μm respectively) produces in Si with the initial temperature of 100 K an increase of both atomic and electronic temperature as shown in Figures 7 – 10, for different sets of $C_e(T_e)$, $K_e(T_e)$.

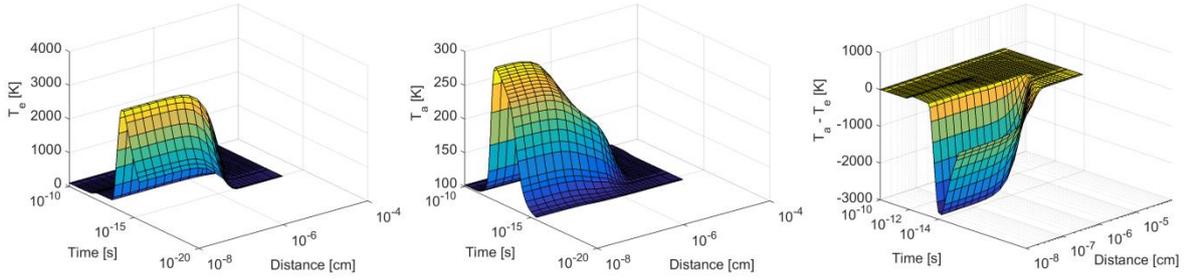

Figure 7: Space and time dependence of $T_e$, $T_a$ and $(T_a - T_e)$ for a Si selfrecoil of 50 keV ($C_e = 3 \cdot 10^{-5} \cdot T_e$, $K_e = 80 \cdot C_e$)

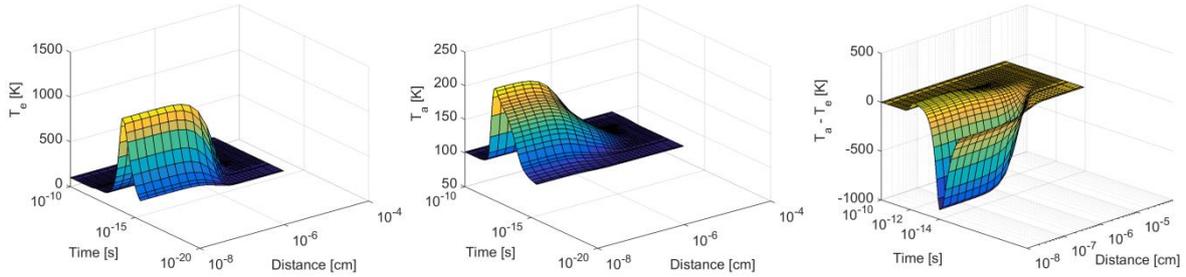

Figure 8: Space and time dependence of $T_e$, $T_a$ and $(T_a - T_e)$ for a Si selfrecoil of 50 keV ($C_e = 0.165$ J/cm$^3$/K, $K_e = 13.2$ W/cm/K)

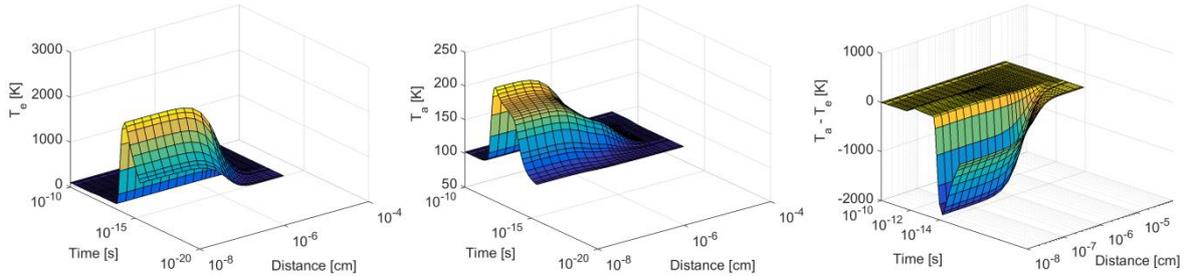

Figure 9: Space and time dependence of $T_e$, $T_a$ and $(T_a - T_e)$ for a Si selfrecoil of 50 keV ($C_e = 3 \cdot 10^{-5} \cdot T_e$, $K_e = 13.2$ W/cm/K)

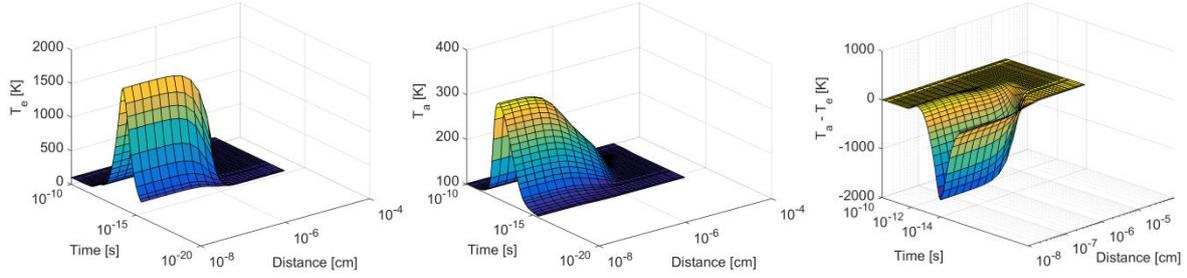

Figure 10: Space and time dependence of $T_e$, $T_a$ and ($T_a - T_e$) for a Si selfrecoil of 50 keV
($C_e$ =0.165 J/cm$^3$/K, $K_e$ =2.4·10$^{-3}$·$T_e$)

Being known that the time constants associated to electronic and atomic processes differ by about two orders of magnitude (see section 3), in the case of lack of coupling the rise of temperature appears distinctly in the two subsystems. The figures evidence the coupling of the two subsystems by the two peaks or a peak and a shoulder in the time dependencies of the two temperatures. The energy flows in both directions: $T_a$ - $T_e$ has a high negative value at very low values of time, and a relatively low positive value which lasts much longer (please note the logarithmic scales both in time and distance).

In Si and Ge, electronic specific heats and conductivities much lower than the atomic ones produce a much higher increase in the electronic temperature for the same deposited energy, and consequently an important difference ($T_e$ - $T_a$), favouring the transfer from the electronic toward the atomic subsystem, especially at high temperatures, where the differences between the parameters of electronic and atomic subsystems become more important.

We would like to mention also that in solving the equations for the thermal spike with initial temperatures below 1 K, the utilization of different types of coupling, i.e. of different exponents *p* as found from experiment, does not modify drastically the solutions of the system of differential equations (1), because the atomic temperature surpasses quickly the limit of 1-2 K which is correlated to the application of the linear energy transfer [30].

### 4.3 Energy transfer between the atomic and electron subsystems due to electron-phonon coupling

During the rise and fall of the thermal spike, the two subsystems (atomic and electronic) exchange energy through the electron-phonon coupling. This energy flows in different points in space, and at different times, either from the atomic toward the electronic subsystems, or reversely, depending on the sign of the difference ($T_a - T_e$).

The energy eventually exchanged between the two subsystems along the whole range of the selfrecoil is found by integrating the exchange term first on the time and in the plane in which the thermal spike develops, thus obtaining the linear energy exchanged in a thin layer perpendicular to the trajectory, which is then integrated on the whole trajectory of the selfrecoil, up to its stop:

$$E_{ex} = \int_0^R dx \int_0^\infty dt \int_0^\infty 2\pi r g (T_a^p - T_e^p) \, dr \tag{13}$$

where $E_{ex}$ represents the energy exchanged between the subsystems, and $R$ the range of the selfrecoil.

In order to evidence the influence of the parameters of the electronic subsystem on the energy transferred, we calculated the linear energy transferred between the atomic and electronic systems of Si, for a selfrecoil of 50 keV, for the same 4 sets of parameters for which the time and space dependencies of the temperatures were calculated (see Figs. 7 – 10), during a single thermal spike developed in a thin layer, perpendicular to its trajectory.

The results are shown in Fig. 11 below. They evidence that the most important influence on the result is given by the electronic specific heat, while the influence of the electronic thermal conductivity is practically negligible.

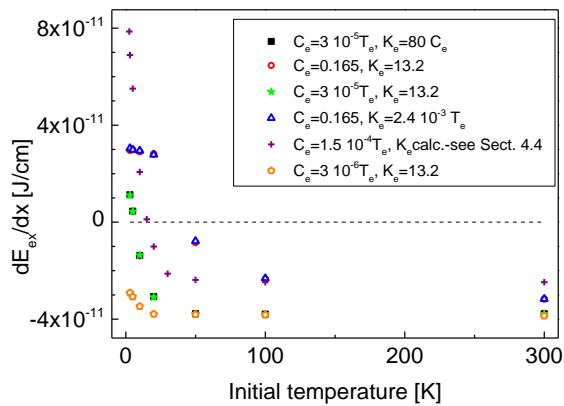

Figure 11: Influence of the parameters of the electronic subsystem on the linear energy transferred between the atomic and electronic subsystems during the thermal spike due to a Si selfrecoil of 50 keV.
$C_e$ is expressed in J/cm$^3$/K and $K_e$ in W/cm/K

One can see that all curves show the same descending trend for the linear exchanged energy with the increase of the temperature of the medium. With the exception of the last set of values, for low initial temperatures (i.e. for detectors working at temperatures below 15 K), the energy is transferred from the atomic toward the electronic subsystem, the value of the energy transferred decreases with the increase of the temperature, and then the direction of transfer is reversed. At temperatures above liquid nitrogen (LN$_2$) a plateau is reached for the transferred energy and the magnitude of $dE_{ex}/dx$ depends weakly on the parameters of the electronic system. In the case of the lowest electronic specific capacity considered, the electronic subsystem transfers always energy to the atomic one for the selfrecoil of 50 keV.

As emphasized at the end of Section 3, the energy exchanged between the two subsystems is calculated based on the hypothesis that all the energy given to each subsystem is used only to rise its temperature, and therefore is able to be exchanged. Therefore, we evaluate the maximum energy exchanged.

## 4.4 Analysis of results for Si and Ge

***Silicon.*** We performed an analysis of the linear energy transfer during the transient processes in Si, with initial temperatures between 2K and 500 K, for selfrecoils of energy between 500 eV and 1 MeV. For the lattice specific heat and thermal conductivity, we used our fit on experimental data (see Section 4.1), while for the electronic subsystem we used the following dependencies: $C_e = 1.5 \times 10^{-4} T_e$ and $K_e(T_e)$ calculated based on formula (10): $K_e = 2.4 \times T_e^{-2.63}$ for $T_e \leq 12K$, $K_e = 0.0035$ for for $12 < T_e \leq 22.4$, $K_e = 18.04 \times T_e^{-2.75}$ for $22.4 < T_e \leq 419.2$ and $K_e = 6 \exp\left(-\frac{6500}{T_e}\right)$ for $T_e > 419.2$ and $g = 1.8 \times 10^{12}$ W/cm³/s. The results, presented in Figure 12, reveal that at very low temperatures always the atomic system transfers energy to the electronic one, and at higher temperatures the reverse is true, and also that the increase of the energy of the selfrecoil produces a small decrease of the temperature of transition between one and the other side of the transfer.

Consequently, in Si cryogenic detectors, for small energy selfrecoils associated with the interaction with WIMPs particles, the transfer of energy from the atomic toward the electronic subsystem is favoured, i.e. the electronic subsystem receives eventually more energy than the one partitioned in agreement with Lindhard curves. At temperatures above about 15 – 20 K, the direction of the transfer is reversed, and the energy in the electronic subsystem is reduced as a consequence of transient phenomena, i.e. the curve describing the partition, $E_{ion}/E_R$ versus $E_R$ is displaced toward lower values.

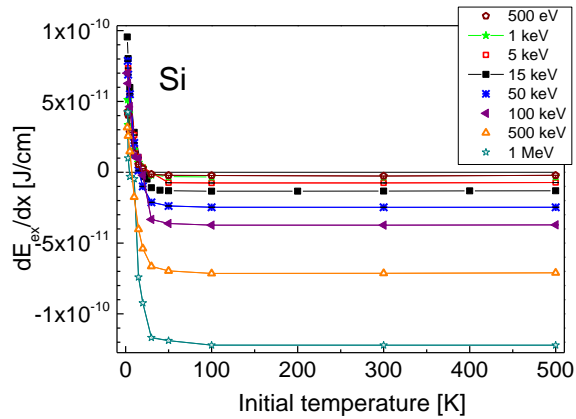

Figure 12: Dependence of the linear energy exchanged during transient thermal processes produced by selfrecoils of different energies, on the initial temperature of Si

***Germanium.*** A similar situation in relation to the parameters of the electronic subsystem, in their dependence on the electronic temperature, and in the dependence of the linear energy transfer on the temperature of the material exists for Ge. Taking $C_e = 2 \times 10^{-5} T_e$, $K_e = 0.13 \cdot T_e^{-2.23}$ for $T_e < 228K$ and $K_e = 0.28 \exp(-2936/T_e)$ at higher electronic temperatures, $g = 2.7 \times 10^{13}$ W/cm³/s, we calculated the linear energy transfer between atomic and electronic subsystems for Ge selfrecoils of different kinetic energies as a function of the initial temperature of the material. The results are presented in Fig. 13. In the case of Ge, the errors in the calculation of the integral (in eq. 13) are important, especially in the temperature region where the linear energy transfer changes sign. The shadowed regions indicate the errors. De ce eroarea la Ge este semnificativa si la Si nu?

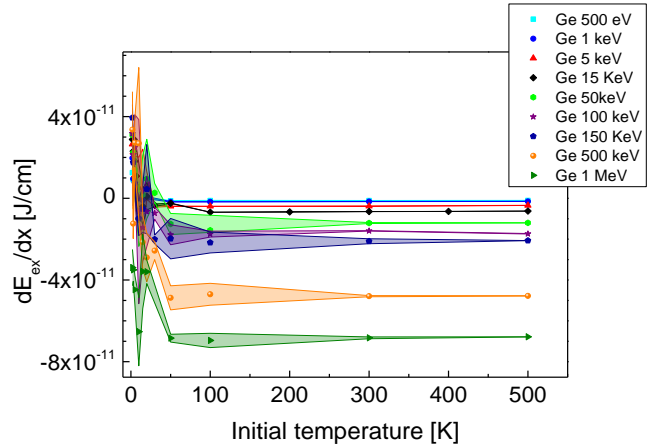

Figure 13: Dependence of the linear energy exchanged during transient thermal processes produced by selfrecoils of different energies, on the initial temperature of Ge

The results obtained for Si and Ge evidence a linear dependence of the plateau value of $dE_{ex}/dx$ on $dE_{ioniz}/dx$, as shown in Figure 14 below. Consequently, for temperatures above $LN_2$, part of the energy of the electronic subsystem is transferred to the atomic one at all energies at the selfrecoil, so that the curves in Figs. 1 and 2 are displaced toward lower values if the correction for the energy exchanged during transient phenomena is considered. Moreover, for Si and Ge are situated on the same curve.

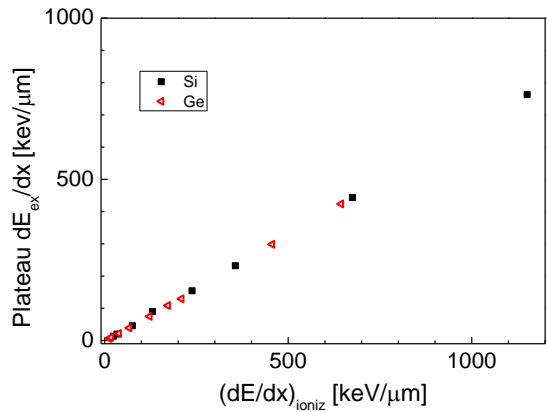

Figure 14: Dependence of the linear exchanged energy on the ionization energy loss for Si and Ge

In contrast to this trend, at very low temperatures, below 15K, for all energies of the selfrecoil, the atomic subsystem transfers energy to the electronic one, so that at these temperatures the Lindhard curves corrected for the energy exchanged during transient phenomena are displaced toward higher values.

An evaluation of the energy transferred between the subsystems during the transient processes for Ge, based on eq. (13), conduces at the result presented in Fig. 15, for low energy selfrecoils. There are two shadowed areas, the first above the Lindhard curve, corresponding to transfer from the nuclear toward the electronic subsystem, which takes place at very low temperatures, and the second one situated below the Lindhard curve, corresponding to transfer from the electronic toward the atomic subsystems in Ge. The upper limit of the first shadowed region corresponds to the maximum transfer atomic-electronic systems (present calculations are for 3 K) and has a maximum at about 6 keV selfrecoil energy. The lower limit of the second shadowed region corresponds to temperatures above LN$_2$, i.e. to the plateau reached in *dE$_{ex}$/dx*. Both these limits are calculated under the mentioned assumption that all the energy in the atomic and electronic sub-systems is stored as heat and is available to be exchanged. The electronic and atomic parameters of Ge and their temperature dependences were the same as specified above.

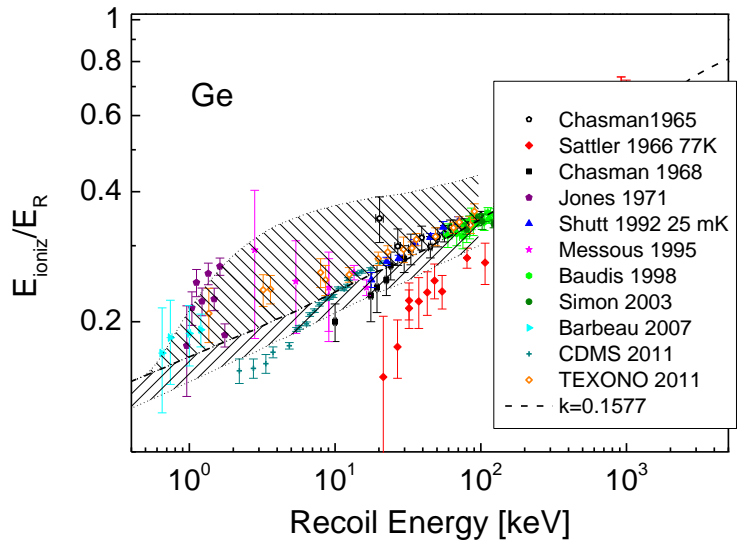

Figure 15: Dependence of energy partition on recoil energy for Ge, with the corrections corresponding to the transferred energy shown as shadowed areas

As can be seen, nearly all the data reported in the literature corresponding to Ge selfrecoils of energy up to 100 keV enter the shadowed areas.

Similar calculations were performed for Si, for the following parameters of the electronic subsystem: $C_e = 3 \times 10^{-6} T_e$ [J/cm$^3$/K], $K_e$ = 13.2 W/cm/K. The result is presented in Fig. 16, with the shadowed regions corresponding to maximum exchanged energy superposed on the compilation of data and on the Lindhard curve for Si.

We would like to emphasise that the upper borders of the shadowed regions depend on the parameters of the electronic system, both in Si and Ge. In the case of Si, using the mentioned dependences Ce(Te) and Ke(Te), the correction to the Lindhard curve corresponding to the energy exchanged at 3 K has a maximum at about 3 keV kinetic energy of the selfrecoil, decreases and passes through zero, changing then sign. This is to be interpreted that, even at 3 K, with the mentioned parameters of the electronic system, there is no exchange from atomic to electronic system in Si.

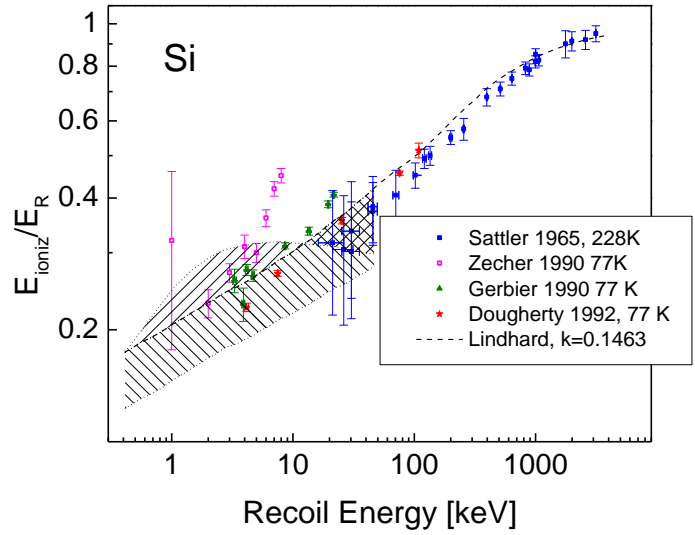

Figure 16: Dependence of energy partition on recoil energy for Si

As can be seen, most of the data reported for low energy Si selfrecoils enter the shadowed area. As remarked in Section 2, a general trend of the data is that they are generally situated under the curve, fact that can be attributed to the energy transferred by the electronic system to the atomic one, at temperatures above LN$_2$ one.

## 5. Summary and conclusions

In this paper, we investigated the influence of the energy exchange between electronic and atomic subsystems during transient thermal processes developed during the slowing down of a selfrecoil in Ge and Si targets, on the partition of its energy. The starting point is Lindhard's theory, the transient processes are treated in the frame of the model of thermal spike, in which the coupling of the subsystems is included as electron-phonon coupling.

In order to estimate the energy exchanged between the subsystems, the knowledge of the temperature dependence of the specific heats and thermal conductivities of the electrons and lattice, as well as of the coupling parameter is necessary.

A review of the data for lattice specific heat and thermal conductivities of Si and Ge is presented, together with a review on the knowledge existent today for the other physical quantities of interest. Due to the lack of consensus on the values and temperature dependences of the electronic specific heats and thermal conductivities, and on the electron-phonon coupling factor, the sensitivity of the temperature distribution in the thermal spike model on these physical quantities was investigated. We found that the most important influence on the result is given by the electronic specific heat, while the influence of the electronic thermal conductivity is a second order effect.

We calculated the energy exchanged between the two subsystems, and found that for both Si and Ge at temperatures higher than LN$_2$, for all recoil energies considered, the linear exchanged energy is temperature independent. More, the values of the plateau of linear exchanged energy have the same linear dependence on the electronic energy loss in both semiconductors analysed (same slope).

In contrast to this, at very low temperatures, below 15 K, for low energy selfrecoils, the atomic subsystem transfers energy to the electronic one.

Consequently, we showed that for low energy of selfrecoils, the corrections to the energy partition curves due to the energy exchange during the transient thermal effects can be divided according to the initial temperature of the target, and have different signs for cryogenic temperatures and temperatures higher than $LN_2$. Experimental data from the literature for Ge and Si fit well this model.

The results are of interest for cryogenic detectors aimed to detect the non-baryonic, non-luminous and non-relativistic dark matter in the Universe, particularly WIMPs.

## Acknowledgements

SL thanks the NIMP Core Programme PN09-450101 for financial support.

## References

[1] L. H. Thomas, The calculation of atomic fields, Proc. Cambridge Philos. Soc. 23 (1927) 542-548.
[2] E. Fermi, Un Metodo Statistico per la Determinazione di alcune Prioprietà dell'Atomo, Rend. Accad. Naz. Lincei 6 (1927) 602-607.
[3] N. Bohr, The penetration of atomic particles through matter, K. Dan. Vidensk. Selsk. Mat. Fys. Medd. 18 (8) (1948) 1-144.
[4] H. Bethe, Zur Theorie des Durchgangs schneller Korpuskularstrahlen durch Materie, Ann. der Physik 397, 325–400, 1930
[5] J. Lindhard, V. Nielsen, M. Scharff, P. Thomsen, Integral equations governing radiation effects, Mat. Fys. Medd. Dan. Vid. Selsk., 33, 10 (1963) 1-42.
[6] C. Chasman, K. W. Jones, and R. A. Ristinen, Measurement of the Energy Loss of Germanium Atoms to Electrons in Germanium at Energies below 100 keV, Phys.Rev. Lett. 15 (1965) 245-248; 15 (1965) 684.
[7] A. R. Sattler, F. L. Vook, and J. L. Palms, Ionization Produced by Energetic Germanium Atoms within a Germanium Lattice, Phys. Rev. 143 (1966) 588-594.
[8] C. Chasman, K. W. Jones, H. W. Kraner, and W. Brandt, Band-Gap Effects in the Stopping of Ge72* Atoms in Germanium, Phys. Rev. Lett. 21 (1968) 1430-1433.
[9] K. W. Jones and H. W. Kraner, Stopping of 1- to 1.8-keV Ge73 Atoms in Germanium, Phys. Rev. C 4 (1971) 125-129.
[10] T. Shutt et al., Measurement of ionization and phonon production by nuclear recoils in a 60 g crystal of germanium at 25 mK, Phys. Rev. Lett. 69 (1992) 3425-3427.
[11] Y. Messous et al., Calibration of a Ge crystal with nuclear recoils for the development of a dark matter detector, Astroparticle Phys. 3 ( 1995) 361-366.
[12] L. Baudis, J. Hellmig, H.V. Klapdor-Kleingrothaus, Y. Ramachers, J.W. Hammer, A. Mayer, High-purity germanium detector ionization pulseshapes of nuclear recoils, -gamma interactions and microphonism, Nucl. Instr. Meth. Phys. Res. A 418 (1998) 348-354.
[13] E. Simon et al., SICANE: a detector array for the measurement of nuclear recoil quenching factors using a monoenergetic neutron beam, Nucl. Instr. Meth. Phys. Res. A 507 (2003) 643–656.
[14] P.S. Barbeau, J.I. Collar and O Tench, Large-Mass Ultra-Low Noise Germanium Detectors: Performance and Applications in Neutrino and Astroparticle Physics, JCAP 09, (2007) 009.
[15] Z. Ahmed et al. (CDMS Coll.), Results from a Low-Energy Analysis of the CDMS II Germanium Data, Phys. Rev. Lett. 106, (2011) 131302.
[16] X. Ruan for CDEX-TEXONO collaboration, "Nuclear recoil quenching factor measurement for HPGe detector", presented at Application of Germanium Detector in fundamental research, Tsinghua University, March 24-26, 2011, https://wwwgerda.mpp.mpg.de/symp/20_Ruan.pdf
[17] A. R. Sattler, Ionization Produced by Energetic Silicon Atoms within a Silicon Lattice, Phys. Rev. 138 (1965) A1815-A1821.
[18] P. Zecher, D. Wang, J. Rapaport, C. J. Martoff, B.A. Young, Energy deposition of energetic silicon atoms within a silicon lattice, Phys. Rev. A 41 (1990) 4058-4061.
[19] G. Gerbier et al., Measurement of the ionization of slow silicon nuclei in silicon for the calibration of a silicon dark-matter detector, Phys. Rev. D 42 (1990) 3211-3215.

[20] B.L. Dougherty, Measurements of ionization produced in silicon crystals by low-energy silicon atoms, Phys. Rev. A 45 (1992) 2104-2107.

[21] A. Mangiarotti, M.I. Lopes, M.L. Benabderrahmane, V. Chepel, A. Lindote, J. Pinto da Cunha, P. Sona, A survey of energy loss calculations for heavy ions between 1 and 100 keV, Nucl. Instr. Meth. Phys. Res. A 580 (2007) 114-117.

[22] I.S. Tilinin, Quasiclassical expression for inelastic energy losses in atomic particle collisions below the Bohr velocity, Phys. Rev. A 51 (1995) 3058-3065.

[23] H. Ascolani and N. R. Arista, Impact-parameter dependence of the electronic energy loss of protons in collisions with atoms, Phys. Rev. A 33 (1986) 2352-2357.

[24] M. Fama, G.H. Lantschner, J.C. Eckardt, N.R. Arista, Angular dependence of the energy loss of ions in solids: Computer simulations and analysis of models, Nucl. Instr. Meth. Phys. Res. B 174 (2001) 16-24.

[25] D. Barker, D.M. Mei, Germanium detector response to nuclear recoils in searching for dark matter, Astropart. Phys. 38 (2012) 1-6.

[26] N. Itoh, D.M. Duffy, S. Khakshouri and A.M. Stoneham, Making tracks: electronic excitation roles in forming swift heavy ion tracks, J. Phys.: Condens. Matter 21 (2009) 474205.

[27] C. Dufour, E. Paumier & M. Toulemonde, A transient thermodynamic model for track formation in amorphous metallic alloys, Radiation Effects and Defects in Solids: Incorporating Plasma Science and Plasma Technology, 126:1-4, (1993) 119-122.

[28] I. Lazanu and S. Lazanu, Transient processes induced by heavy projectiles in silicon, Nucl. Instrum. Meth. B 268 (2010) 2241-2245.

[29] S. Lazanu, I. Lazanu and G. Ciobanu, Modelling the transient processes produced under heavy particle irradiation, Nucl. Instrum. Meth. B 269 (2011) 498-503 [arXiv:1011.6611].

[30] I. Lazanu, S. Lazanu, Transient thermal effects in solid noble gases as materials for the detection of Dark Matter, JCAP 07 (2011) 013 [ArXiv ePrint: 1103.1841].

[31] I. Lazanu, ML Ciurea, S. Lazanu, Analysis of defect formation in semiconductor cryogenic bolometric detectors created by heavy dark matter, Astropart. Phys. 44 (2013) 9–14.

[32] C. Kittel, Introduction to Solid State Physics, 7th Ed., Wiley, (1996).

[33] M. Blackman, The theory of the specific heat of solids, Rep. Prog. Phys. 8 (1941) 11-30.

[34] N. Pearlman and P.H. Keesom, The Atomic Heat of Silicon below 100°K, Phys. Rev. 88 (1952) 398-405.

[35] P.H. Keesom and N. Pearlman, The Atomic Heat of Germanium below 4°K, Phys. Rev. 91 (1953) 1347-1353.

[36] P. Flubacher, A. J. Leadbetter, and J. A. Morrison, The heat capacity of pure silicon and germanium and properties of their vibrational frequency spectra, Philos. Mag. 4 (1959) 273-294.

[37] A. S Okhotin, A. S. Pushkarskii, and V. V. Gorbachev, Thermophysical Properties of Semiconductors, Moscow, "Atom" Publ. House 1972, from the database of Ioffe Physicotechnical Institute at
http://www.ioffe.rssi.ru/SVA/NSM/Semicond/Si/thermal.html

[38] S Adachi, Handbook on physical properties of semiconductors, Vol 1, Kluwer Academic publishers, 204, pag. 49-50.

[39] U.Piesbergen, Die durchschnittlichen Atomwärmen der AmBv-Halbleiter AlSb, GaAs, GaSb, InP, InAs, InSb und die Atomwärme des Elements Germanium zwischen 12 und 273 °K, Z. Naturforshung 18a (1963) 141-147.

[40] T.O. Niinikoski, A. Rijllart, A. Alessandrello, E. Fiorini, A. Giuliani, Heat Capacity of a Silicon Calorimeter at Low Temperatures Measured by Alpha-Particles Europhys Lett 1 (1986) 499-504.

[41] S. Wagner, M. Lakner and H.v Löhneysen, Specific heat of Si:(P,B) at low temperatures Phys. Rev. B 55 (1997) 4219-4224.

[42] G. D. Tsibidis, M. Barberoglou, P.A. Loukakos, E. Stratakis, C. Fotakis, Dynamics of ripple formation on silicon surfaces by ultrashort laser pulses in subablation conditions Phys. Rev. B 86 (2012) 115316.

[43] G.G. de la Cruz and Yu.G. Gurevich, Time-dependent heat diffusion in semiconductors by electrons and phonons, Phys. Rev. B 58 (1998) 7768-7773.

[44] A. Kamarou, W. Wesch, E. Wendler, A. Undisz, and M. Rettenmayr, Swift heavy ion irradiation of InP: Thermal spike modeling of track formation, Phys. Rev. B 73 (2006) 184107.

[45] A. Kamarou, W. Wesch, E. Wendler, A. Undisz, and M. Rettenmayr, Radiation damage formation in InP, InSb, GaAs, GaP, Ge, and Si due to fast ions, Phys. Rev. B 78 (2008) 054111.

[46] A. Chettah, H. Kucal, Z.G. Wang, M. Kac, A. Meftah, M. Toulemonde, Behavior of crystalline silicon under huge electronic excitations: A transient thermal spike description, Nucl. Instr. Meth. Phys. Res. B 267 (2009) 2719-2724.

[47] M. Toulemonde, W. Assmann, C. Dufour, A. Meftah, F. Studer and C. Trautmann, Experimental Phenomena and Thermal Spike Model Description of Ion Tracks in Amorphisable Inorganic Insulators, Mat. Fys. Medd. 52 (2006) 263-292.

[48] Z.G. Wang, Ch. Dufour, E. Paumier and M. Toulemonde, On the Se sensitivity of metals under swift-heavy-ion irradiation: a transient thermal process, J. Phys.: Condens. matter 7 (1995) 2525-2526.

[49] Ch Dufour, V Khomenkov, G Rizza and M Toulemonde, Ion-matter interaction: the three-dimensional version of the thermal spike model. Application to nanoparticle irradiation with swift heavy ions, J. Phys. D: Appl. Phys. 45 (2012) 065302.

[50] CA Bryant and PH Keesom, Low-Temperature Specific Heat of Germanium, Phys. Rev. 124 (1961) 698-700.

[51] E Aubourg et al., Measurement of Electron-Phonon Decoupling Time in Neutron-Transmutation Doped Germanium at 20 mK, J. Low Temp. Phys. 93 (1993) 289-294.

[52] N Wang, FC Wellstood, EE Haller, J Beeman, Electrical and thermal properties of neutron-transmutation-doped Ge at 20 mK, Phys. Rev. B 41 (1990) 3761-3748.

[53] E. Olivieri, M. Barucci, J. Beeman, L. Risegari, G. Ventura, Excess Heat Capacity in NTD Ge Thermistors J. Low Temp. Phys. 143 (2006) 153-162.

[54] R. E. Peierls, Quantum Theory of Solids (Oxford U. P., London, 1955).

[55] J. M. Ziman, Electrons and Phonons (Oxford U. P., London, 1960).

[56] P. Carruthers, Theory of Thermal Conductivity of Solids at Low Temperatures, Rev. Mod. Phys. 33 (1961) 92-138.

[57] J. R. Drabble and H. J. Goldsmid, Thermal Conduction in Semiconductors (Pergamon, Oxford, England,1961).

[58] J Callaway, Model for Lattice Thermal Conductivity at Low Temperatures, Phys. Rev. 113 (1959) 1046-1051.

[59] H.B.G. Casimir, Note on the Conduction of Heat in Crystals Physica 5 (1938) 495-500.

[60] M Kazan, G Guisbiers, S Pereira, M R Correia, P Masri, A Bruyant, S Volz, and P. Royer, Thermal conductivity of silicon bulk and nanowires: Effects of isotopic composition, phonon confinement, and surface roughness J. Appl. Phys. 107 (2010) 083503.

[61] G. A. Slack and S. Galginaitis, Thermal Conductivity and Phonon Scattering by Magnetic Impurities in CdTe, Phys. Rev. 133 (1964) A253-A268.

[62] I. Pomeranchuk, On the Thermal Conductivity of Dielectrics, Phys. Rev. 60 (1941) 820-821.

[63] C J Glassbrenner, G A Slack, Thermal Conductivity of Silicon and Germanium from 3° K to the Melting Point, Phys. Rev. 134 (1964) 1058-1069.

[64] M.G. Holland, Analysis of Lattice Thermal Conductivity, Phys. Rev. 132 (1963) 2461–2471.

[65] G A Slack, Thermal conductivity of pure and impure silicon, silicon carbide and diamond, J. Appl. Phys. 35 (1964) 3460-3465.

[66] G A Slack, C J Glassbrenner, Thermal conductivity of germanium from 3°K to 1020° K, Phys. Rev. 120 (1960) 782-789.

[67] H J Albany, M Vandevyver, Low temperature thermal conductivity of fast neutron irradiated silicon and germanium, J. Appl. Phys. 38 (1967) 425-430.

[68] J R Carruthers, T H Geballe, H M Rosenborg, and J M Ziman, The thermal conductivity of germanium and silicon between 2 and 300 K, Proc. Roy . Soc. (London) 238A (1957) 502-514

[69] J A Carruthers, J F Cochran and K Mendelssohn, Thermal conductivity of p-type germanium between 0.2° and 4°K, Cryogenics 2 (1962) 160-166.

[70] TH Geballe and G W Hull Isotopic and other types of thermal resistance in germanium, Phys. Rev. 110 (1958) 773-775.

[71] C.Y.Ho, R.W. Powell, E.Liley, Thermal conductivity of the elements: a comprehensive review, J. Phys. & Chemical Reference Data vol. 3, 1974, supplement No. 1 (published by the American Chemical Soc and the AIP for NIST)

[72] R.W. Ure, Jr. Proc of the IEEE/AIAA thermoelectric Specialists Conf, Washington, D.C. May 17, 1966, paper No. 11

[73] D. A. Damon and R. W. Ure, Jr. and G. Gersi, The Lattice thermal conductivity of Bismuth Telluride and some Bismuth Telluride - Bismuth Sellenide Alloys, in Thermal Conductivity, Proc of the Seventh Conference on Thermal Conductivity, National Bureau of Standards, Nov. 13 -16 , 1967, eds. D.R. Flynn, B.A. Peavy, Jr. pag.111-122.

[74] B.I. Davydov and M. Shmushkevich, Uspekhi Fiz. Nauk 24 (1940) 21.

[75] P.J. Price, Ambipolar thermodiffusion of electrons and holes in semiconductors, Philosophical Magazine Series 7, 46 (1955) 1252-1260.

[76] G Grimvall, The Electron–Phonon Interaction in Metals (Oxford: North-Holland,1981).

[77] C. Dufour, A. Audouard, F. Beuneu, J. Dural, J.P. Girard, A. Hairie, M. Levalois, E. Paumier, M. Toulemonde, A high resistivity phase induced by swift heavy ion irradiation of Bi: a probe for thermal spike damage, J. Phys. Condens. Matter. 5 (1993) 4573–4584.

[78] C.P. Flynn, R.S. Averback, Electron-phonon interactions in energetic displacement cascades, Phys. Rev. B 38 (1988) 7118-7120.

[79] M.W. Finnis, P. Agnew, A.J.E. Foreman, Thermal excitation of electrons in energetic displacement cascades, Phys. Rev. B 44 (1991) 567-574.

[80] M. I. Kaganov, I. M. Lifshitz, and L. V. Tanatarov, Relaxation between electrons and the crystalline lattice, Sov. Phys. JETP 4, (1957) 173.

[81] C P Race, D R Mason, M W Finnis, W M C Foulkes, A P Horsfield, and A P Sutton, The treatment of electronic excitations in atomistic models of radiation damage in metals, Rep. Progr. Phys. 73 (2010) 116501.

[82] J. K. Chen, W. P. Latham, and J. E. Beraun, J. Laser Appl. Vol 17, no. 1 (2005) 63: Anisimov, in [S. I. Anisimov, B. L. Kapeliovich, and T. L. Perel'man, "Electron emission from metal surfaces exposed to ultrashort laser pulses," Sov. Phys.JETP 39, 375–377 s1974d]

[83] M. Y. Reizer, Electron-phonon relaxation in pure metals and superconductors at very low temperatures. Phys. Rev. B 40 (1989) 5411 – 5416.

[84] P. Kivinen, PhD Thesis, "Electrical and thermal transport properties of semiconductor and metal structures at low temperature", University of Jyvaskyla, 2005, http://www.jyu.fi/static/fysiikka/vaitoskirjat/2005/pasi_kivinen.pdf

[85] M Toulemonde, WJ. Weber, G Li, V Shutthanandan, P Kluth, Y Tang, Y Wang, and Y Zhang, Synergy of nuclear and electronic energy losses in ion-irradiation processes: The case of vitreous silicon dioxide, Phys. Rev. B 83 (2011) 054106.

[86] R. Horn, J.P. Harrison, Electron-Phonon Coupling in Bolometers Used for Cryogenic Particle Detection, J. Low Temp. Phys. 133 (2003) 291–311.

[87] R.J. Zieve, D.E. Prober, R.G. Wheeler, Low-temperature electron-phonon interaction in Si MOSFETs, Phys. Rev. 57 (1998) 2443–2446.

[88] M. Prunnila, J. Ahopelto, A.M. Savin, P.P. Kivinen, J.P. Pekola, A.J. Manninen, Electron-phonon coupling in degenerate silicon-on-insulator film probed using superconducting Schottky junctions, Physica E 13 (2002) 773–776.

[89] R. Leoni, B. Buonomo, G. Castellano, F. Mattioli, G. Torrioli, L. Gaspare, F. Evangelisti, Doped silicon and NIS junctions for bolometer applications, Nucl. Instr. and Meth. Phys. Res A 520 (2004) 44–47.

[90] K. Suzuki and N.Mikoshiba, Low-Temperature Thermal Conductivity of p-Type Ge and Si, Phys. Rev. B 3 (1971) 2550 – 2556.

[91] T Sota and K Suzuki, Acoustic properties of heavily doped many-valley semiconductors in the weak-localization regime. Phys. Rev. B 33 (1986) 8458 – 8467.

[92] P. Kivinen, A. Savin, M. Zgirski, P. Torma, J. Pekola, Electron–phonon heat transport and electronic thermal conductivity in heavily doped silicon-on-insulator film, J. Appl. Phys. 94 (2003) 3201-3205.

[93] J. Clerouin, Atelier modélisation et procédés lasers ultra-brefs, Carry Le Rouet Mars 2010 http://reseau-femto.cnrs.fr/IMG/pdf/CLEROUIN.pdf